\documentclass[seceq]{ptptex}
\newcommand{\be}{\begin{equation}}
\newcommand{\ee}{\end{equation}}
\newcommand{\ba}{\begin{eqnarray}}
\newcommand{\ea}{\end{eqnarray}}

\newcommand{\hh}{\, ,\hspace{0.5cm}}
\newcommand{\hhh}{\, ,\hspace{0.2cm}}

\newcommand{\hook}{\raisebox{-0.35ex}{\makebox[0.6em][r]
{\scriptsize $-$}}\hspace{-0.15em}\raisebox{0.25ex}{\makebox[0.4em][l]{\tiny
 $|$}}}
\newcommand{\eq}[1]{(\ref{#1})}

\newcommand{\n}[1]{\label{#1}}
\newcommand{\bs}[1]{{\boldsymbol{#1}}}

\newcommand{\bsn}[2]{{\boldsymbol{#1}}_{#2}}
\newcommand{\bsnn}[2]{{\boldsymbol{#1}}_{\bar{#2}}}

\newcommand{\ve}{\varepsilon}
\newcommand{\bse}[2]{{\boldsymbol{#1}}^{#2}}
\newcommand{\bsee}[2]{{\boldsymbol{#1}}^{\bar{#2}}}

\title{%        %You can use \\ for explicit line-break
Hidden Symmetries of Higher-Dimensional 
Black Hole Spacetimes%
}

\author{%       %Use \scshape  for the family name
Valeri P.\textsc{Frolov}%
}

\inst{%     %Affiliation, neglected when [addenda] or [errata]
Theoretical Physics Institute, University of Alberta, Edmonton,
Alberta, Canada T6G 2G7
}

\abst{
The paper contains a brief review of recent results on
hidden symmetries in higher dimensional black hole spacetimes. We
show how the existence of a principal CKY tensor (that is a closed
non-degenerate conformal  Killing-Yano 2-form) allows one to generate
a `tower' of Killing-Yano and Killing tensors responcible for hidden
symmetries. These symmetries imply complete integrability of geodesic
equations and the complete separation of variables in the
Hamilton-Jacobi, Klein-Gordon and Dirac equations in the general
Kerr-NUT-(A)dS metrics. 
}

\begin{document}

\maketitle

\section{Introduction}

The concept of symmetries is one of the most powerful tools of modern
theoretical physics. Noether's theorem relates continuous symmetries 
to conservation laws. The most fundamental of them are connected with
the symmetries of the background spacetime. In a curved spacetime 
continuous isometries are generated by Killing vector fields. For
each of the Killing vector fields there exists a conserved quantity.
For example, for a geodesic particle motion this conserved quantity is
a projection of the particle momentum on the Killing vector. Besides these
symmetries, which are `naturally' connected with the spacetime
isometries, there exist hidden symmetries, generated by either
symmetric or antisymmetric tensor fields. These objects are connected
with conserved quantities which are higher than first order in
momentum.

A well known example is the famous Kerr spacetime. It was
demonstrated by Carter \cite{Car:68a} that the Hamilton-Jacobi
equation for a particle motion in the Kerr metric allows a separation
of variables and the geodesic equations are completely integrable.
The natural conservation laws connected with Kerr metric symmetries,
which is stationay and axisymmetric, are not sufficient to explain
this `miracle'. Really, spacetime symmetries are `responcible' for two
integrals of motion, the energy and the azimuthal component of the
angular momentum. This together with the conservation of $\bs{p}^2$
gives only 3 integrals of motion. Carter \cite{Car:68a} constructed the
fourth required integral of motion, which is quadratic in momentum
and is connected with the Killing tensor $K_{ab}$ \cite{WP}. Penrose
and Floyd \cite{PF} showed that this Killing tensor is a `square' of
an atisymmetric Killing-Yano tensor \cite{Yano}. 

In many aspects a Killing-Yano tensor is more fundamental than a
Killing tensor. Namely, its `square'   is always Killing tensor, but
the opposite is not generally true (see, e.g., \cite{Ferrando}).  In 
$4$--dimensional spacetime, as it was shown by Collinson
\cite{Coll74},  if a vacuum solution of the Einstein equations allows
a non--degenerate Killing-Yano tensor it is of the type D.  All the
vacuum type D solutions were obtained by Kinnersley \cite{Kinn}.
Demianski and Francaviglia \cite{DeFr} showed that in the absence  of
the  acceleration these solutions admit Killing and Killing-Yano
tensors.  It should be  also mentioned that if a spacetime admits a
non--degenerate Killing-Yano tensor it always has at least one
Killing vector  \cite{Dietz}.

Recently a lot of interest focuses on higher dimensional black hole
solutions. In the widely discussed models with large extra dimensions
it is assumed that one or more additional spatial dimensions are
present. In such models one expects mini black hole production in the
high energy collisions of particles.  Mini black holes can serve as a
probe of the extra dimensions. At the same time their interaction
with the brane, representing our physical world, can give the 
information about the brane properties. If a black hole is much
smaller that the size of extra dimension and the brane tension can be
neglected, its metric is an asymptotically flat or (A)dS solution of
higher dimensional Einstein equations. The most general known black
hole solution, which besides the mass and rotation parameters, admits
aslo NUT parameters and the cosmological constant was obtained 
\cite{pope}. Recently it was achieved a remarcable progress in study
the properties of such black holes. Namely, it was demonstrated that
in many respects these metrics are similar to the 4-dimensional
Kerr-NUT-(A)dS metric. They possess wide enough symmetry to make it
possible complete integrability of the geodesic motion equations and
the complete separation of variables in the Hamilton-Jacobi,
Klein-Gordan and Dirac equations. Main technical tools for obtaining
these results is connected with the existence of hidden symmetries.
This paper contains a brief review of these results.

\section{Killing-Yano and Killing tensors}

Let us consider a $D$-dimensional spacetime with a metric $\bs{g}$.
In order to cover both cases of odd and even dimensions
we write the spacetime dimension $D$ in the form $D=2n+\varepsilon$,
where $\ve=0$ ($\ve=1$) for the even (odd) dimensional case. 
One says that 
the spacetime  has a symmetry generated by the
vector field $\xi^{a}$ if  this vector obeys the {\em
Killing equation} 
\be 
\xi_{(a;b)}=0\, . 
\ee 
For a geodesic motion
of a particle in such a curved spacetime the quantity
$p^{a}\xi_{a}$, where $p^{a}$ is its momentum, remains constant
along the particle trajectory. Similarly, for a null geodesic,
$p^{a}\xi_{a}$ is conserved provided $\xi^{a}$ is a {\em conformal
Killing vector} obeying the equation
\be
\xi_{(a;b)}=\tilde{\xi}g_{ab}\hh \tilde{\xi}=D^{-1}\xi^{b}_{\ \ ;b}\, .
\ee

There exist natural generalizations of this symmetry which result in
the conservation laws for objects higher than of the first power in
the momentum ${\bs{p}}$. In such a case one says that the spacetime
possesses a {\em hidden symmetry}. 

A {\em symmetric} tensor $K_{a_1a_2 \ldots a_p}=
K_{(a_1a_2 \ldots a_p)}$ is called a
{conformal Killing tensor} if it obeys the equation
\be\n{CK}
K_{(a_1a_2 \ldots a_p;b)}=g_{b (a_1}\tilde{K}_{a_2 \ldots
a_p)}
\, .
\ee
As in the case of a conformal Killing vector, the tensor
$\tilde{\bs{K}}$ is  determined by traciong the both sides of
the equation \eq{CK}. If $\tilde{\bs{K}}$ vanishes, the tensor
$\bs{K}$ is called a {\em Killing tensor}. In a presence of the
Killing tensor ${\bs K}$ the conserved quantity for a geodesic motion
is $K_{a_1a_2 \ldots a_p}p^{a_1}p^{a_2}\ldots p^{a_p}$.
For null geodesics this quamtity is conserved not only for the Killing
tensor, but also for a conformal Killing tensor.

An {\em antisymmetric} generalization of the Killing vector is known
as a {\em Killing-Yano tensor}. An antisymmetric tensor 
$h_{a_1a_2 \ldots a_p}=h_{[a_1a_2 \ldots a_p]}$ is called
a {\em conformal Killing-Yano tensor} (or, briefly, CKY tensor) if it
obeys the following equation \cite{cari}
\be\n{CKY}
\nabla_{(a_1}h_{a_2)a_3 \ldots a_{p+1}}=
g_{a_1a_2}\tilde{h}_{a_3 \ldots a_{p+1}}-
(p-1)g_{[a_3(a_1}\tilde{h}_{a_2) \ldots a_{p+1}]}\, .
\ee 
By tracing the both sides of this equation one obtains the following
expression for $\tilde{\bs{h}}$
\be
\tilde{h}_{a_2 a_3 \ldots a_{p}}={1\over D-p+1}\nabla^{a_1}h_{a_1a_2
\ldots a_p}\, .
\ee
In the case when $\tilde{\bs{h}}=0$ one has a {\em 
Killing-Yano tensor}, or, briefly, KY tensor. If $f_{a_1a_2 \ldots
a_p}$ is a Killing-Yano tensor than $K_{ab}=f_{a a_2 \ldots
a_p}f_{b}^{\ \ a_2 \ldots
a_p}$ is a Killing tensor.

Let  $h_{ab}$ be a CKY tensor then the vector
\be\n{prime}
\xi^{(0)a}={1\over D-1}\nabla_b h^{ab}
\ee
obeys the following equation \cite{jez}
\be
\xi^{(0)}_{\ (a;b)}=-{1\over D-2}R_{\sigma (\mu}h_{\nu)}^{\,\,\,\,\sigma}\,
.
\ee
Thus, in an Einstein space, that is when
$R_{ab}=\Lambda g_{ab}$, $\bs{\xi}^{(0)}$ is the Killing vector.

\section{Killing-Yano equations in terms of differential forms}

The CKY tensors are forms and operation with them are greatly
simplified if one uses the "language" of differential forms. 
We just remind some of the relations we use in the present paper.
If $\bs{\alpha}_p$ and $\bs{\beta}_q$ are $p$- and $q$-forms,
respectively, the external derivative ($d$) of their external product
($\wedge$) obeys a relation
\be\n{dd}
d(\bs{\alpha}_p\wedge \bs{\beta}_q)=
d\bs{\alpha}_p\wedge \bs{\beta}_q+
(-1)^p\bs{\alpha}_p\wedge d\bs{\beta}_q\, .
\ee
A
Hodge dual $*\bs{\alpha}_p$ of the $p$-form $\bs{\alpha}_p$ is
$(D-p)$-form defined as
\be
* \bs{\alpha}_p \Leftrightarrow e_{a_1 \ldots a_{D-p}}^{\quad\quad
a_{D-p+1}\ldots a_D}\omega_{a_{D-p+1}\ldots a_D}\, ,
\ee
where $e_{a_1 \ldots a_D}$ is a totally anti-symmetric tensor.
The exterior {\em co-derivative}
$\delta $ is defined as follows
\be
\delta \bs{\alpha}_p=(-1)^p\epsilon_p *d*\bs{\alpha}_p\hhh
\epsilon_p=(-1)^{p(D-p)}{\mbox{det}(g)\over |\mbox{det}(g)|}\, .
\ee
One also has $* *\bs{\alpha}_p=\epsilon_p\bs{\alpha}_p$.

If $\{ \bs{e}_a\}$ is a basis of vectors, then dual basis of 1-formes
$\bs{\omega}^a$ is defined by the relations
$\bs{\omega}^a(\bs{e}_b)=\delta^a_b$. We denote
$\eta_{ab}=g(\bs{e}_a,\bs{e}_b)$ and by $\eta^{ab}$ the inverse
matrix. Then the operations with the indices enumerating the basic
vectors and forms are performed by using these matrices. In
particular, $\bs{e}^a=\eta^{ab}\bs{e}_b$, and so on. We denote a
covariant derivative along the vector $\bs{e}_a$ by
$\nabla_a=\nabla_{\bs{e}_a}$. One has
\be
d=\bs{\omega}^a\wedge \nabla_a\hhh
\delta=-\bs{e}^a\hook \nabla_a\, .
\ee
In the tensor notations the `hook' operator applied to a $p$-form
$\bs{\alpha}_p$ corresponds to a contraction 
\be
\bs{X}\hook \bs{\alpha}_p \Leftrightarrow X^{a_1} \alpha_{a_1 a_2
\ldots a_p}\, .
\ee
For a given vector $\bs{X}$ one  defines $\bs{X}^{\flat}$   as a
corresponding 1-form with the components
$(X^{\flat})_{a}=g_{ab}X^{b}$. 
In particular, one has $\eta^{ab}(\bs{e}_b)^{\flat}=\bs{\omega}^a$.
We refer to \cite{sten,kress} where  these and many other useful
relations can be found.

The definition \eq{CKY} of the CKY tensor $\bs{h}$ (which is a
$p$-form)  is equivalent to the following equation (see e.g.
\cite{kress})  
\be\n{CKYf}
\nabla_\bs{X} \bs{h}={1\over p+1} \bs{X}\hook d\bs{h}
-{1\over D-p+1}\bs{X}^{\flat}\wedge
\delta \bs{h}\, .
\ee 
If $\delta \bs{h}=0$ this is an equation for the Killing-Yano tensor.

Using a relation
\be
\bs{X}\hook *\bs{\omega}=*(\bs{\omega}\wedge \bs{X}^{\flat})\, 
\ee
it is easy to show that \eq{CKYf} implies
\be
\nabla_\bs{X}(* \bs{h})={1\over p_*+1} \bs{X}\hook d(*\bs{h})
-{1\over D-p_*+1}\bs{X}^{\flat}\wedge
\delta(*\bs{h})\hhh p_*=D-p\, .
\ee
It means that a Hodge dual $*\bs{h}$ of a CKY tensor $\bs{h}$ is again
a CKY tensor. Moreover, if the CKY is closed, $d\bs{h}=0$, then its
dual $(D-p)$-form $f=*\bs{h}$ is a Killing-Yano tensor.

The following result \cite{KKPF} plays an important role in the
construction of the hidden symmetry objects in higher dimensional
spacetimes. 

{\bf Proposition.} {\em If $\bs{h}^{(1)}$ and $\bs{h}^{(2)}$ are two closed CKY
tensors their external product $\bs{h}=\bs{h}^{(1)}\wedge
\bs{h}^{(2)}$ is also a closed CKY tensor.}

To prove this proposition we first notice that the equation \eq{dd}
implies that the form $\bs{h}$ is closed. Suppose now
that a $p$-form $\bs{\alpha}_p$ obeys a relation
\be\n{eq0}
\nabla_{\bs{X}}\bs{\alpha}_p=\bs{X}^{\flat}\wedge \bs{\gamma}\, ,
\ee 
then
\be\n{eq1}
\bs{\gamma}=-{1\over D-p+1}\delta\bs{\alpha}_p\, .
\ee
Really
\ba\n{eq2}
\delta\bs{\alpha}_p&=&-\bs{e}^a \hook \nabla_a \bs{\alpha}_p=
- \bs{e}^a \hook (\bs{\omega}_a \wedge \bs{\gamma})\nonumber\\
&=&-(\bs{e}^a\hook \bs{\omega}_a)\wedge
\bs{\gamma}-\bs{\omega}_a\wedge(\bs{e}^a\hook
\bs{\gamma})=-(D-p+1)\bs{\gamma}\, .
\ea
We used the relations
\be
\bs{e}^a\hook \bs{\omega}_a=D\hhh
\bs{\omega}_a\wedge (\bs{e}^a\hook \bs{\gamma})=(p-1)\bs{\gamma}\, .
\ee
The relation \eq{eq1} follows from \eq{eq2}. 

The second step in the proof of the Proposition is to show that if
$\bs{\alpha}_p$ and $\bs{\beta}_q$ are two closed CKY tensors then
\be
\nabla_{\bs{X}}(\bs{\alpha}_p\wedge
\bs{\beta}_q)=\bs{X}^{\flat}\wedge\bs{\gamma}_{p+q-1}\, .
\ee
Really, one has
\ba\n{eq3}
&&\nabla_{\bs{X}}(\bs{\alpha}_p\wedge
\bs{\beta}_q)=
\nabla_{\bs{X}}\bs{\alpha}_p\wedge
\bs{\beta}_q+
\bs{\alpha}_p\wedge
\nabla_{\bs{X}}\bs{\beta}_q\\
&&=-{1\over D-p+1}(\bs{X}^{\flat}\wedge \delta \bs{\alpha}_p)\wedge
\bs{\beta}_q
-{1\over D-q+1}
\bs{\alpha}_p\wedge (\bs{X}^{\flat}\wedge \delta
\bs{\beta}_q)
= \bs{X}^{\flat}\wedge \bs{\gamma}_{p+q-1}\, ,\nonumber\\
&& \bs{\gamma}_{p+q-1}=-{1\over D-p+1}\delta \bs{\alpha}_p\wedge
\bs{\beta}_q
-{(-1)^p\over D-q+1}
\bs{\alpha}_p\wedge \delta
\bs{\beta}_q\, .
\ea
Combining the relations \eq{eq0}-\eq{eq3} one arrives to the result
given in the Proposition.

\section{Principal CKY tensors and  `towers' of Killing and
Killing-Yano tensors}

Let us
consider now a special case which is important for applications.
Namely we assume that a spacetime allows a 2-form $\bs{h}$ which is a
{\em closed conformal Killing-Yano tensor}. Such a 2-form can be
written, at least locally, as $\bs{h}=d\bs{b}$.  We also assume that
the 2-form $h$ is {\em non-degenerate}, that is its rank is $2n$. We
call such an object a {\em principal CKY tensor}. 

It is easy to show that $S_{ab}=h_{ac}h^c_{\ \ b}$ is a symmetric
tensor, and its eigen-values $x^2$,
\be\n{problem}
S^a_{\ \ b}v^{b}=-x^2 v^a\, ,
\ee 
are real and non-negative. Using a modified Gram-Schmidt procedure it
is possible to show that there exists such an orthonormal basis in
which the operator $h^a_{\ \ b}$ has the following form
\be\n{dar}
\bs{h}=\mbox{diag}(0,\ldots,0,\bs{\Lambda}_1,\ldots,\bs{\Lambda}_p)\, ,
\ee
where $\bs{\Lambda}_i$ are matrices of the form
\be
\bs{\Lambda}_i=\left(   
\begin{array}{cc}
0 & -x_i \bs{I}_i\\
x_i \bs{I}_i& 0
\end{array}
\right)\, ,
\ee
and $\bs{I}_i$, are unit matrices. Such a basis is known as the {\em
Darboux basis} (see e.g. \cite{pras}). Its elements are unit
eigen-vectors of the problem \eq{problem}.

For a non-degenerate 2-form $\bs{h}$ the number of zeros
in the Darboux decomposition \eq{dar} consides with $\ve$. If all the
eigen-values $x$ in \eq{problem} are different (we denote them
$x_{\mu}$, $\mu=1,\ldots,n$), the matrices
$\bs{\Lambda}_i$ are 2-dimensional. Denote the vectors of the Darboux
basis by $\bsn{e}{\mu}$ and $\bsnn{e}{\mu}\equiv \bsn{e}{n+\mu}$,
where $\mu=1,\ldots,n$. In the odd dimensional spacetime we also have
an additional basic vector $\bsn{e}{0}$ (an eigen-vector of
\eq{problem} with $x=0$). Orthonormal vecors $\bsn{e}{\mu}$ and
$\bsnn{e}{\mu}$ span a 2-dimensional plane of eigen-vectors of
\eq{problem} with the same eigen-value $x_{\mu}$. We denote by
$\bse{\omega}{\mu}$ and $\bsee{\omega}{\mu}\equiv \bse{\omega}{n+\mu}$ (and
$\bse{\omega}{0}$ if $\ve=1$) the dual basis of 1-forms.
The metric $g_{ab}$ and the principal CKY tensor $\bs{h}$ in this basis take
the form
\ba
g_{ab}&=&\sum_{\mu=1}^n (\bse{\omega}{\mu}_{a}\bse{\omega}{\mu}_{b}+
\bsee{\omega}{\mu}_{a}\bsee{\omega}{\mu}_{b})+\ve \bse{\omega}{0}_a\bse{\omega}{0}_b\,
,\n{gab}\\
\bs{h}&=&\sum_{\mu=1}^n x_{\mu} \bse{\omega}{\mu}\wedge \bsee{\omega}{\mu}\,
.\n{hab}
\ea

According to the
Proposition of the previous section, the principal CKY tensor
generates a set (`tower') of new closed CKY tensors
\be
\bs{h}^{(j)}=\bs{h}^{\wedge j}=\underbrace{\bs{h}\wedge \ldots \wedge
\bs{h}}_{\mbox{\tiny{total of $j$ factors}}}\, .
\ee 
$\bs{h}^{(j)}$ is a $2j$ form. In particular for $j=1$
$\bs{h}^{(1)}=\bs{h}$. Since $\bs{h}$ is non-degenerate, one has a
set of $n$ non-vanishing closed CKY tensors. In the even dimensional
spacetime  $\bs{h}^{(n)}$ is proportional to the totally antisymmetric
tensor.

Each $2j$-form 
$\bs{h}^{(j)}$ determines a $(D-2j)$-form of the Killing-Yano tensors 
\be
\bs{f}^{(j)}=*\bs{h}^{(j)}\, .
\ee
In its turn, these tensors determine the Killing tensors
$\bs{K}^{(j)}$ \cite{KKPF}
\be\n{KK}
K^{(j)}_{ab}={1\over (D-2j-1)!(j!)^2} f^{(j)}_{\, \, \, \, \, a c_1\ldots c_{D-2j-1}}
f_{\, \, \, \, \, b}^{(j) \, \,  c_1\ldots c_{D-2j-1}}\, .
\ee
A choice of the coefficient in the definition \eq{KK} is a matter of 
convenience. It is convenient to include the metric $\bs{g}$, which
is a trivial Killing tensor, as an element $\bs{K}^{(0)}$ of the
`tower' of the Killing tensors. The total number of elements of the `extended
tower' is $n$ \footnote{For example, in 5D spacetime where $n=2$,
this `tower' contains only one non-trivial
Killing tensor. For the 5D rotating black hole solution this Killing
tensor was first found in \cite{FSa,FSb} by using the Carter's method
of separation of variables in the Hamilton-Jacobi equation.}

\section{Hidden symmetries of the higher dimensional Kerr-NUT-(A)dS
spacetimes}

The most general known higher dimensional solution describing rotating black
holes with NUT parameters in an asymptotically (Anti) deSitter
spacetime (Kerr-NUT-(A)dS metric) was obtained in \cite{pope}. This
metric has the form \eq{gab} where
\be \n{bas_e}
\bs{\omega}^\mu=\frac{dx_\mu}{\sqrt{Q_\mu}}\hhh
\bs{\omega}^{\bar{\mu}}=\sqrt{Q_\mu}\sum_{k=0}^{n-1}A^{(k)}_{\mu} d\psi_k\,
\hhh
\bs{\omega}^{0} = (c/A^{(n)})^{1/2}
\sum_{j=0}^nA^{(j)}d\psi_j\, .
\ee
Here 
\be\n{QU}
Q_{\mu}=X_{\mu}/U_{\mu}\hh
U_{\mu}\equiv\prod_{\nu\ne\mu}(x_{\nu}^2-x_{\mu}^2)\,.
\ee
and the coefficients $A^{(k)}_{\mu}$ and $A^{(j)}$ are polynomial
functions of coordinates $x_{\nu}$ determined by the following
expansions (their explicit form can be found in \cite{pope})
\be\n{part1}
\prod_{\nu=1}^{n}(1+tx_{\nu}^2)=\sum_{j=0}^n t^j A^{(j)}\hhh
(1+tx_{\mu}^2)^{-1}\prod_{\nu=1}^{n}(1+tx_{\nu}^2)=\sum_{k=0}^{n-1}
t^k A^{(k)}_{\mu}\, .
\ee
The coefficients $X_{\mu}$ are functions of one variable, $x_{\mu}$,
which for the Kerr-NUT-(A)dS metric are 
\be\n{XX}
X_\mu =\sum_{k=\ve}^{n}\, c_{k}\, x_\mu^{2k}+ b_\mu\, x_\mu^{1-\ve}
+\ve\frac{(-1)^n c}{x_\mu^2}\, .
\ee 
This metric is of the Petrov type {\bf D} \cite{col}.
The total number of constants which enter the solution is $2n+1$: 
$\ve$ constants $c$, $n+1-\ve$ constants $c_k$ and $n$ constants
$b_{\mu}$. The form of the metric is invariant under a 1-parameter 
scaling coordinate transformations, thus a total number of
independent parameters is $D-\ve$. These parameters are related to
the cosmological constant, mass, angular momenta, and NUT parameters.
One of these parameters can be used to define a scale, while the
other $D-1-\ve$ parameters can be made dimensionless \cite{pope}. 
This solution may be considered as a higher dimensional
generalization of 4-dimensional Ker-NUT-(A)dS  solution obtained by
Carter \cite{Car:68b}. Moreover, the coordinates used in the metric
\eq{gab}, \eq{bas_e}-\eq{XX} are higher dimensional analogue of the
Debever-Carter coordinates \cite{deb,Car:68c} .

The curvature for the general metric \eq{bas_e}-\eq{part1} with
arbitrary functions $X_{\mu}(x_{\mu})$ was calculated in \cite{hhoy}. For a
special choice \eq{XX} this metric is a solution of the higher
dimensional Einstein equations
\be\n{Eeq}
R_{ab}=\Lambda g_{ab}\hhh
\Lambda=(D-1)(-1)^n c_n\, .
\ee

It was shown in \cite{KF,FK} that this spacetime possesses a principal
CKY tensor $\bs{h}$ which has the form \eq{hab} and its potential
$\bs{b}$, $\bs{h}=d\bs{b}$, is \footnote{In fact, this potential
generates a principal CKY tensor for a general form of the metric
\eq{gab},\eq{bas_e}--\eq{part1} with arbitrary functions
$X_{\mu}(x_{\mu})$.}
\be
\bs{b}={1\over 2}\sum_{k=0}^{n-1} A^{(k+1)} d\psi_k\, .
\ee
The Killing tensors associated with the principal CKY tensor for the
Kerr-NUT-(A)dS spacetime can be written as follows \cite{KKPF}
\be
{K}^{(j)}_{ab}=\sum_{\mu=1}^n A_{\mu}^{(j)} 
(\bse{\omega}{\mu}_{a}\bse{\omega}{\mu}_{b}+
\bsee{\omega}{\mu}_{a}\bsee{\omega}{\mu}_{b})
+\ve c^{-1} A^{(n)} A^{(j)}\bse{\omega}{0}_a\bse{\omega}{0}_b\, .
\ee

We call the Killing vector $\bs{\xi}^{(0)}$  generated by the principal CKY
tensor $\bs{h}$ (see \eq{prime}) a {\em primary Killing vector}. In
the Kerr-NUT-(A)dS spacetime the primary Killing vector is
$\bs{\xi}^{(0)}=\partial_{\psi_0}$.  Besides the primary Killing vector, 
this spacetime has $n-1$ additional Killing vectors  $\bs{\xi}^{(j)}$
\cite{KKPF}
\be\n{kvs}
\xi^{(j)a}=K^{(j)}\,\!^a_{\ b}\xi^b\hhh
\xi^{(j)a}\partial_{a}=\partial_{\psi_j}\hhh
j=1,\dots,n-1\, .
\ee
In odd dimensions the last Killing vector is given by
the ${n}$-th Killing--Yano tensor  $\bs{f}^{(n)}$, which in the
Kerr-NUT-(A)dS spacetime  turns out to be $\partial_{\psi_n}$.
The total number of these  Killing vectors is $n+\ve$. For a geodesic
motion they give $n+\ve$ linear in momentum integrals of motion. The
`extended tower' of the Killing tensors $\bs{K}^{(j)}$ ($j=0,\ldots n-1$) gives
$n$ additional integrals of motion, which are quadratic in the
momentum. Thus the total number of conserved quantities for a geodesic
motion is $2n+\ve$, that is it coincides with the number of the
spacetime dimensions $D$. It is possible to show that these integrals of
motion are independent and in involution, so that the geodesic motion in the
Kerr-NUT-(A)dS spacetime is completely integrable
\cite{PKVK,PKVKb,hoy1}.
Moreover, it was shown recently \cite{sekr} that the following
operators
\ba\n{LLL}
\bs{L}_{(k)}&=&-i\xi^{(k)a}\partial_{a}\hhh (k=0,\ldots,n+\ve-1)\, ,\\
\bs{K}_{(j)}&=&-{1\over\sqrt{|g|}}
\partial_{a}[\sqrt{|g|}K^{(j)\,\! a  b}\partial_{b}]\hhh (j=0,\ldots,n-1)\, ,
\n{KKK}
\ea
determined by a principal CKY tensor, are mutually commutative.

It should be emphasized that the coordinates in the metric 
\eq{bas_e}--\eq{part1} have a well defined geometrical meaning. The
`essential' coordinates $x_{\mu}$ are connected with eigen-values of
the principal CKY tensor $\bs{h}$ (see \eq{hab}), while the Killing
coordinates $\psi_{j}$ are defined by the Killing vectors generated by
the principal CKY tensor. Namely this invariant definition of the
coordinates in the metric \eq{bas_e}--\eq{part1} makes it so convenient
for calculations.

The existence of a principal CKY tensor imposes non-trivial
restrictions on the geometry of the spacetime. Namely, the following
result was  proved in~\cite{hoy2}. Let $\bs{h}$ be a principal CKY
tensor and $\bs{{\xi}^{(0)}}$ be its primary Killing vector. Then if
\be\n{cond}
{\cal L}_{\bs{\xi}^{(0)}} \bs{h}=0\, ,
\ee
then the only solution of the the Einstein equations with the cosmological
constant \eq{Eeq} is the Kerr-NUT-(A)dS spacetime. (Here ${\cal L}_{\bs{u}}$ is a Lie
derivative along the vector $\bs{u}$.)

\section{Hidden symmetries and separation of variables}

The massive scalar field equation
\be\n{seq}
\Box \Phi-m^2\Phi=0\, ,
\ee
in the Kerr-NUT-(A)dS metric allows a complete separation of variables
\cite{FrKrKu}. Namely a solution can be decompose into modes
\be\n{mod}
\Phi=\prod_{\mu=1}^n R_{\mu}(x_{\mu})\prod_{k=0}^{n+\ve-1}e^{i\Psi_k
\psi_k}\, .
\ee
Substitution of \eq{mod} into the equation \eq{seq} results in the
following {\em second order ordinary differential equations} for functions
$R_{\mu}(x_{\mu})$
\be
(X_{\mu}R'_{\mu})'+\ve{X_{\mu}\over
x_{\mu}}R'_{\mu}+\left(V_{\mu}-{W_{\mu}^2\over X_{\mu}}\right)R_{\mu}=0\, .
\ee
Here
\be\n{fun}
W_{\mu}=\sum_{k=0}^{n+\ve-1}\Psi_k(-x_{\mu}^2)^{n-1-k}\hhh
V_{\mu}=\sum_{k=0}^{n+\ve-1}\kappa_k(-x_{\mu}^2)^{n-1-k}\, .
\ee
Here $\kappa_0=-m^2$
and for $\ve=1$ we put $\kappa_n=\Psi_n^2/c$. The parameters
$\kappa_k$ ($k=1,\ldots n+\ve-1$)
are separation constants. Using \eq{KKK} one has $\bs{K}_{(0)}=-\Box$.
Since all the operators \eq{LLL}--\eq{KKK} commute with one another,
their common eigen-values can be used to specify the modes. It is possible
to show \cite{sekr} that the eigen-vectors of these commuting operators are
the modes \eq{mod} and one has
\be
\bs{L}_{(k)}\Phi=\Psi_k\Phi\hhh
\bs{K}_{(j)}\Phi=\kappa_j \Phi\, .
\ee

Similarly, the Hamilton-Jacobi equation for geodesic motion 
\be
{\partial S\over \partial \lambda}+g^{ab}\partial_a S \partial_a S=0\,
,
\ee
in the Kerr-NUT-(A)dS spacetime allows a complete separation of
variables \cite{FrKrKu}
\be
S=-w\lambda+\sum_{k=0}^{n+\ve-1}\Psi_k\psi_k+\sum_{\mu=1}^n
S_{\mu}(x_{\mu})\, .
\ee
The functions $S_{\mu}$ obey the first order ordinary differential
equations
\be
{S'_{\mu}}^2={V_{\mu}\over X_{\mu}}-{W_{\mu}^2\over X_{\mu}^2}\, ,
\ee
where the functions $V_{\mu}$ and $W_{\mu}$ are defined in \eq{fun}.

Recently it was shown that the massive Dirac equation in the
Kerr-NUT-(A)dS spacetime also allows the separation of variables
\cite{oy}. It was also proved that the stationary test string
equations in the  Kerr-NUT-(A)dS spacetime are completely integrable
\cite{KF_string} .

\section{Conclusions}

The Kerr-NUT-(A)dS metric is the most general known solution
describing higher dimensional rotating black hole spacetimes with NUT
parameters in an asymptotically (Anti) de Sitter spacetime
background. It possesses, what we called, a principal CKY tensor
$\bs{h}$ which determines the hidden symmetries of this spacetime.
This 2-form $\bs{h}$ generates a `tower' of Killing-Yano and Killing
tensors, which make it possible a complete integrability of geodesic
equations and separability of the Hamilton-Jacobi, Klein-Gordon and
Dirac equations. Moreover, if a higher dimensional solution of the
Einstein equations with the cosmological constant  allows a principal
CKY tensor obeying \eq{cond}, it coincides with the Kerr-NUT-(A)dS
metric. These remarkable properties of higher dimensional black hole
solutions resemble the well known `miraculuous' properties of the
Kerr spacetime. Does this analogy goes further? Are all higher
dimensional solutions with the principle CKY tensor of Petrov type
{\bf D}? Can the higher spin massless field equations be decoupled
and do they allow separation of variables? These are interesting but
still open problems.

\section*{Acknowledgements}

The author thanks  the Yukawa Institute for Theoretical Physics at
Kyoto University, where this work was partially done during the
scientific  program on "Gravity and Cosmology 2007, and Professor
Misao Sasaki for the hospitality. He also is greatful to the Natural
Sciences and Engineering Research Council of Canada and the Killam
Trust for the financial support. 

%\appendix
%\section{First Appendix} %Empty argument \section{} yields `Appendix'. 
%
%\section{Second Appendix}

\end{document}